\documentclass[journal]{vgtc}                

\ifpdf
  \pdfoutput=1\relax                   
  \usepackage{graphicx}                
  \DeclareGraphicsExtensions{.pdf,.png,.jpg,.jpeg} 
\else
  \usepackage{graphicx}                
  \DeclareGraphicsExtensions{.eps}     
\fi%

\graphicspath{{figures/}{pictures/}{images/}{./}} 

\usepackage{microtype}                 
\PassOptionsToPackage{warn}{textcomp}  
\usepackage{textcomp}                  
\usepackage{mathptmx}                  
\usepackage{times}                     
\usepackage{cite}                      
\usepackage{tabu}                      
\usepackage{booktabs}                  

\onlineid{0}

\vgtccategory{Research}
\vgtcpapertype{please specify}

\title{ARbis Pictus: A Study of Vocabulary Learning with\\ Augmented Reality}

\author{Adam Ibrahim, Brandon Huynh, \textit{Student Member, IEEE}, Jonathan Downey, \\
Tobias H\"{o}llerer, \textit{Senior Member, IEEE},  Dorothy Chun, and John O'Donovan}
\authorfooter{
\item
 Adam Ibrahim is with the Department of Computer Science at the University of California,  Santa Barbara. E-mail: ai@cs.ucsb.edu.
\item
 Brandon Huynh is with the Department of Computer Science at the University of California,  Santa Barbara. E-mail: bhuynh@cs.ucsb.edu.
\item
 Jonathan Downey is with the Department of Education at the University of California, 
 Santa Barbara. E-mail: jonathandowney@ucsb.edu.
\item
 Tobias H\"{o}llerer is with the Department of Computer Science at the University of California,  Santa Barbara. E-mail: holl@cs.ucsb.edu.
\item
 Dorothy Chun is with the Department of Education at the University of California, 
 Santa Barbara. E-mail: dchun@education.ucsb.edu.
 \item
 John O'Donovann is with the Department of Computer Science at the University of California,  Santa Barbara. E-mail: jod@cs.ucsb.edu.
}

\shortauthortitle{Ibrahim \MakeLowercase{\textit{et al.}}: ARbis Pictus: A study of Vocabulary Learning with Augmented Reality}

\abstract{
We conducted a fundamental user study to assess potential benefits of AR technology for immersive vocabulary learning. With the idea that AR systems will soon be able to label real-world objects in any language in real time, our within-subjects (N=52) lab-based study explores the effect of such an AR vocabulary prompter on participants learning nouns in an unfamiliar foreign language, compared to a traditional flashcard-based learning approach. Our results show that the immersive AR experience of learning with virtual labels on real-world objects is both more effective and more enjoyable for the majority of participants, compared to flashcards. Specifically, when participants learned through augmented reality, they scored significantly better on both same-day and 4-day delayed productive recall tests than when they learned using the flashcard method. We believe this result is an indication of the strong potential for language learning in augmented reality, particularly because of the improvement shown in sustained recall compared to the traditional approach.  
} 


 \CCScatlist{ 
   \CCScat{H.5.1}{Information Interfaces and Presentation}{Multimedia Information Systems}%
 {Artificial, augmented, and virtual realities};
   \CCScat{H.5.2}{Information Interfaces and Presentation}{User Interfaces}{Training, help, and documentation};
 }
 
\begin{document}


\firstsection{Introduction}

\maketitle

This paper addresses the problem of facilitating and understanding the process of language learning in immersive, augmented reality (AR) environments. Recent heavy investment in AR technology by industry leaders such as Google, Microsoft, Facebook and Apple is an indication that both device technology and content for this modality will improve rapidly over the coming years. Looking forward, we believe that AR can have significant impact on the way we learn foreign or technical languages, processes and workflows, for example, by creating new personalized learning opportunities in a physical space that is modeled, processed and labeled by automated machine learning (ML) classifiers, assisted by human users. These augmented learning environments can include annotations on real objects, placement of virtual objects, or interactions between either type to describe complex processes. Thus, without significant extra user effort, a future "always-on"  AR system could seamlessly provide a user with the foreign-language terms describing objects (or later possibly even processes) in their own physical environments, enabling casual reminders and incidental learning of vocabulary. 

Such learned language skills will still be in effect and valuable when the AR device is not in use. This is in contrast to using automatic translation services, which can also strongly benefit from AR technology but which require being online and may not actually help in {\em learning} a language if relied upon without reflection, much like unconsidered reliance on online navigation services might not improve navigation skills, and in fact can lead to inferior spatial knowledge \cite{Ishikawa2008}.

AR devices will eventually become affordable and portable enough to be commonly used in day-to-day tasks. In this setting, learning can occur passively as people interact with objects and processes in their environments that are annotated to support personalized learning objectives.
\par

\begin{figure}
  \centering
  \includegraphics[width=3in]{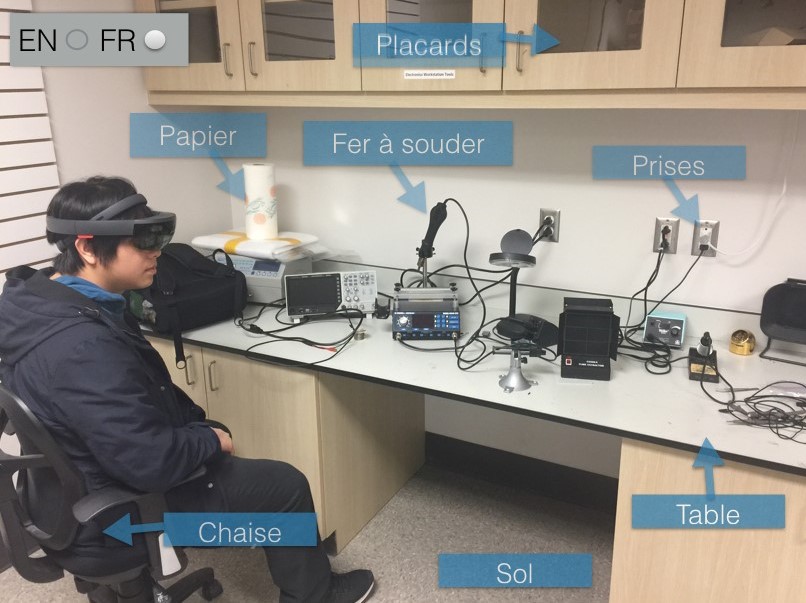}
  \caption{An illustrative mock up of a language learner using the ARbis Pictus system. Note that the user will only see annotations in an approximately $30^\circ\times 17^\circ$ field of view, due to current constraints with the HoloLens.}~\label{fig:overview}
\end{figure}    

Our project towards this vision is called `ARbis Pictus'', named after Orbis Sensualium Pictus (Visible World in Pictures), one of the first widely used children's picture books, written by Comenius and first published in 1658. To study the benefits of using AR for memorization tasks such as used in language learning, we conducted a foundational user study to evaluate the impact of learning simple noun terms in a foreign language with augmented reality labeling using a purpose-built AR object labeling tool. While the scenario given above for the medium-to-longer-term vision includes the potential for casual incidental learning, this first study examines active intentional learning efforts. We focus on the following three research questions in our user study:

\begin{itemize}
    \item RQ 1: When learning vocabulary (or individual lexical items) in an unknown second language, is there a benefit of learning with AR over a traditional flashcard-based method? 
    \item RQ 2: In the above setting, how does productive recognition and recall vary after some time has passed?
    \item RQ 3: How do users \textit{perceive} the language learning experience in Augmented Reality compared to traditional flashcards? 
\end{itemize}

In the process of answering these research questions, we make the following contributions, results and insights. To carry out our study, we designed and implemented a system that supports foreign language vocabulary learning with augmented reality and with traditional flashcards. We then designed and implemented a user experiment to evaluate the impact of AR-based learning for foreign language vocabulary.
Key findings include 1) better recall (7\%, p$<$0.05) for AR learning compared to traditional flashcards; 2) an increased advantage (21\%, p$<$0.01) for AR in productive recall four days after the initial test, compared to traditional flashcards; 3) Qualitative survey and interview data shows that participants believe that AR is effective and enjoyable for language learning.  During the study, attention and gaze data was collected through the AR device, through an eye tracker, and through click interactions.  We also describe an early-stage analysis of this data, and how it reveals learning patterns in each modality.

\section{Background}
Our study aims to show that there is a measurable benefit to learning vocabulary through AR labeling of real world objects.  Before we discuss the design, we briefly describe existing work on multimedia learning and on AR in education settings. 

\subsection{Multimedia Learning}
Our framework is motivated by Mayer et al.'s cognitive theory of multimedia learning (CTML) \cite{mayer2005cambridge}\cite{mayer2011applying}, one of the most compelling learning theories in the field of Educational Technology. The theory posits, first, that there are two separate channels (auditory and visual) for processing information, second, that learners have limited cognitive resources, and third, that learning entails active cognitive processes of filtering, selecting, organizing, and integrating information. The CTML predicts, based on extensive empirical evidence, that people learn better from a combination of words and pictures than from words alone\cite{mayer1994whom}. In the field of Second Language Acquisition (SLA), studies using the CTML as their theoretical basis have shown that when unknown vocabulary words are annotated with both text (translations) and pictures (still images or videos), they are learned and retained better in post tests than words annotated with text alone \cite{chun1996effects}\cite{plass1998supporting}\cite{yoshii2006}. A second principle of the CTML is that people learn better when corresponding words and pictures are presented near rather than far from each other on the page or screen, as the easy integration of verbal and visual information causes less cognitive load on working memory, thereby facilitating learning \cite{moreno1999cognitive}. SLA research has found that simultaneous display of multimedia information leads to better performance on vocabulary tests than an interactive display \cite{turk2014effects}. Our approach extends the CTML by simultaneously displaying information next to physical objects, allowing learners to further integrate spatial information of the object and its surroundings.

A recent study by Culbertson et al.
in \cite{culbertson2016social} describes an online 3D game to teach people Japanese.
Their approach used situated learning theory, and they found
excellent feedback on engagement. Specifically, people were
learning 8 words in 40 min on average. Experts who already
knew some Japanese were the most engaged with the system.
Learning results from that study informed the design and complexity of the learning tasks in our experiment. The broader vision for our ARbis Pictus system, including personalized learning and real-time object recognition was influenced by work by Cai et al. in \cite{cai2014wait}, which found that we can leverage the small waiting times in everyday life to teach people a foreign language, e.g. while chatting with a friend electronically.

\subsection{Virtual and Augmented Reality in Education}
The use of Augmented Reality for second language learning is in its infancy \cite{scrivner2016augmented}\cite{godwin2016augmented}, and there are only a small number of studies that link AR and second language learning. For example, in \cite{liu2016analyzing}, Liu et al. describe an augmented reality game that allows learners to collaborate in English language learning tasks. They find that the AR approach increases engagement in the learning process. In contrast, our experiment is an evaluation of the effects of immersive AR on lexical learning, using simple noun terms only, analogous more to a traditional flashcard-based learning method. The benefits and shortcomings of flashcards are well documented in the second language learning literature \cite{nakata2011computer}. In this study, we employ this method as a simple benchmark, purposely chosen to minimize effects of user interactions, and to expose the impact of immersion in AR. Grasset et al.\cite{grasset2007mixed} and Scrivner et al. \cite{scrivner2016augmented} have studied AR textbooks in the classroom. Their approach differs from ours in that we use minimal virtual objects (labels only), but incorporate physical objects in the real world as a pedagogical aid, including their spatial positioning in the augmented scene. Godwin-Jones\cite{godwin2016augmented} provides a review of AR in language learning, focusing on popular games such as Pokemon Go! and general AR devices and techniques, but doesn't discuss formal evaluations of AR for second language learning. Going beyond simple learning of lexical terms, the European Digital Kitchen project \cite{seedhouse2014european} incorporates process-based learning with AR to support language learning.  They apply a marker-based tracking solution to place item labels in the environment to help users prepare recipes, including actions such as stirring, chopping or dicing, for example. Dunleavy et al. \cite{dunleavy2014augmented} discuss AR and situated learning theory.  They claim that immersion helps in the learning process, but also warn about the dangers of increased cognitive overload that comes with AR use.  In our experimental design, we consider this advice and allow ample time for familiarization with the AR device to reduce both cognitive overload resulting from the unfamiliar modality, and other novelty effects. 

 \par
 
 AR has been shown in first experiments to support better memorization of items \cite{Rosello2016}, \cite{Fujimoto2012a}, making use of spatial organization and the memory palace technique. Our study is in line with these promising results and shows that there can be a distinct benefit of AR for vocabulary learning, comparing with tried-and-true flashcard-based approaches.
 
 Another benefit of AR is that it brings an element of gamification to the learning task, making it particularly suitable for children to learn with. There have been several interactive games involving AR for learning in a variety of situations.  Costabile et al.\cite{costabile2008explore} discuss an AR application for teaching history and archaeology. Like \cite{dunleavy2014augmented}, they hypothesized that engagement would be increased with AR compared to more traditional displays. However, the results found that a traditional paper method was both faster and more accurate than AR for the learning task.  Another notable example is Yannier et al.'s study \cite{yannier2015learning} on the pedagogy of basic physical concepts such as balance, using blocks.  In their study, AR outperformed benchmarks by about a 5-fold increase, and was reported as far more enjoyable. A similar, but much earlier approach that applied AR to collaboration and learning was Kaufman's work \cite{kaufmann2003mathematics} on teaching geometry to high-school level kids. An updated version of this system was applied to mobile AR devices by Schmalstieg et al. in  \cite{schmalstieg2007experiences}.  Now that we have described relevant related work that has informed our experimental design and setup, we can proceed with details of our designs. This will be followed with a discussion of results.

\begin{table*}[]
\centering
\begin{tabular}{@{}lcccccc@{}}
\textit{Order} & \multicolumn{6}{l}{\textit{Device used | word subgroup(s) seen during each learning phase}} \\ \midrule
I     & AR - A1 & AR - A1, A2 & AR - A1, A2, A3 & FC - B1 & FC - B1, B2 & FC - B1, B2, B3 \\    
II    & AR - B1 & AR - B1, B2 & AR - B1, B2, B3 & FC - A1 & FC - A1, A2 & FC - A1, A2, A3 \\    
III   & FC - A1 & FC - A1, A2 & FC - A1, A2, A3 & AR - B1 & AR - B1, B2 & AR - B1, B2, B3 \\    
IV    & FC - B1 & FC - B1, B2 & FC - B1, B2, B3 & AR - A1 & AR - A1, A2 & AR - A1, A2, A3 \\
\end{tabular}
\caption{Table of conditions and balancing across the six learning phases.  AR shows the augmented reality conditions and FC represents flashcards. A and B are distinct term groups for the within-subject design, and the group number indicates one of the subgroups of 5 words.}
\label{tab:ordering}
\end{table*}

\section{Experimental design}
52 participants (33 females, 19 males, mean age of 21, SD of 3.8) took part in a within-subject study. Learning modality and word groups were systematically varied, resulting in a 2x2 counterbalanced design.  Participants were recruited through a paid pool at a university, and were a mix of students and non-students. 

In terms of ethnic background, 16 participants self-identified as White or Caucasian, 14 as Asian, 9 as Hispanic or Latina/Latino, 3 as Asian and Caucasian, 4 as Latino/Latina and Caucasian, 3 as African American, 1 as African American and Mexican, 1 as Middle Eastern, and 1 as Mixed without further elaboration.
No participant had high proficiency in three or more languages. 19 participants reported being natively or fully professionally proficient in a second language. Of the other 33 participants, 4 reported speaking more than one language with intermediate proficiency, and 24 with some elementary proficiency. 16 people reported having some kind of proficiency in three or more languages. 

English was by far the most commonly well-spoken language, with 48 participants reporting "Native or Multi-lingual Proficiency" in it, and 4 participants "Full Professional Proficiency". The second most commonly well-spoken language was Spanish, with 30 participants reporting having any kind of fluency in it.  The third most spoken language group consisted of Chinese languages, as reported by 9 participants. No participant reported any kind of proficiency in Basque, the language of choice for our experiment.

30 Basque words were divided into two groups of 15, called A and B, further divided into fixed subgroups of 5 referred to as A1, A2, A3 and B1, B2 and B3. Each subject saw one of the two word groups on one of the devices, and the other group on the other device. In total, 13 people saw the word group A in AR first, 13 word group B in AR first, 13 word group A with the flashcards first, and 13 word group B with the flashcards first, as described in Table \ref{tab:ordering}.

After answering a set of background questions, participants were told what the objects were in English and started using one of the devices after the instructors informed them about the learning tasks and the specifics of the tests. On the AR device, the participants first undertook a training task where they could take as much time as they wanted to set up the device, get used to the controls and reduce the novelty aspect of it while interacting with virtual objects. Before using the flashcards, an eye-tracker was calibrated for each participant. Then, the participants moved on to the learning task, which consisted of 3 learning phases and 4 tests (3 productive recognition tests, 1 productive recall test) per device. In the first learning phase, the participants had 90 seconds to learn the 5 words of the first subgroup of one of the word groups on a given device. After a distraction task, they took a productive recognition test. Afterwards, they undertook a second learning phase on the same device, and had 90 seconds to learn the 5 new words from the second subgroup of the same word group, along with the 5 previous words. Following a distraction task, they took a recognition test on the 5 new words. They then had a third learning phase on the same device during which they saw for 90 seconds the 5 words of the last subgroup of the selected word group, alongside the 10 previous ones. After a distraction and a recognition test on the 5 words from the last subgroup, they took a productive recall test on all 15 words from the word group chosen. They then had another, similar set of 3 learning phases and 4 tests on the other device using the other word group, as illustrated in Table \ref{tab:ordering}. The AR learning task, flashcard learning task and tests took place in 3 different rooms to avoid potential biases. 

At the conclusion of the learning task, the participants answered a questionnaire on how efficient and engaging they perceived each device. A short interview allowed us to gather more feedback on their preferences. Four days after the learning phases, the participants were asked to take again the same 8 tests they took the day of the study to assess long-term recall. 32 users agreed to take the tests.

Every participant was compensated \$10, and the study lasted a total of 40 to 65 minutes for every user (with most of the variance due to the AR training phase's flexible length).

\section{Experimental setup}
Our experiment setup consisted of two modalities. An AR learning tool and a traditional flashcard tool, implemented as a browser-based web application.  We now present the details of both, along with the learning and distraction tasks that were completed by each participant. 

\subsection{Flashcards}

\begin{figure}
  \centering
  \includegraphics[width=3.3in]{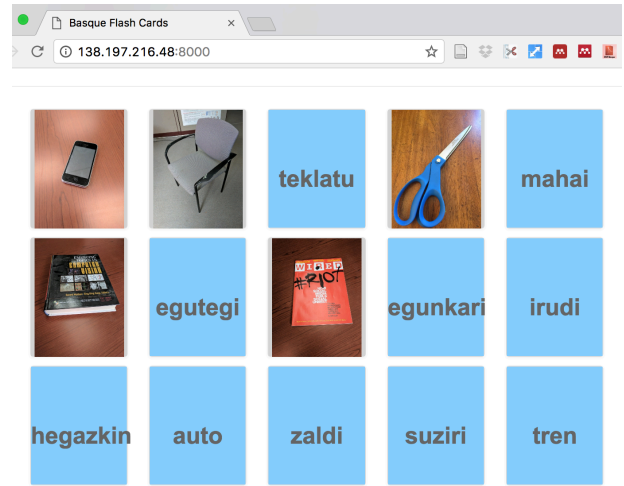}
  \caption{Screen shot of the web-based flashcard application that was used in the study. }~\label{fig:flash}
\end{figure}

The flashcard modality was designed as a web application emulating traditional physical flashcards, running on a desktop computer that the user interacted with using a mouse (see Figure \ref{fig:flash}). After entering a user ID and one of the combinations of word subgroups seen in Table \ref{tab:ordering}, the instructor let the participants interact with 1, 2 or 3 rows of 5 flashcards, all visible on a single page, with each flashcard consisting of a word in the foreign language on the back and an image of the corresponding object on the front. The images used were pictures of the real objects used in the AR condition. A recording of the word being pronounced was automatically played through speakers every time the user clicked on the back of a flashcard. The same recording of the Basque word being spoken by a human (male) was used in both modalities.  Clicks were logged during every phase to track possible learning strategies. Additionally, an eye tracker was calibrated before the learning task with the flashcards to track the participants' gaze during the learning phases.

\subsection{Augmented Reality}
\begin{figure}
\centering
  \includegraphics[width=0.9\columnwidth]{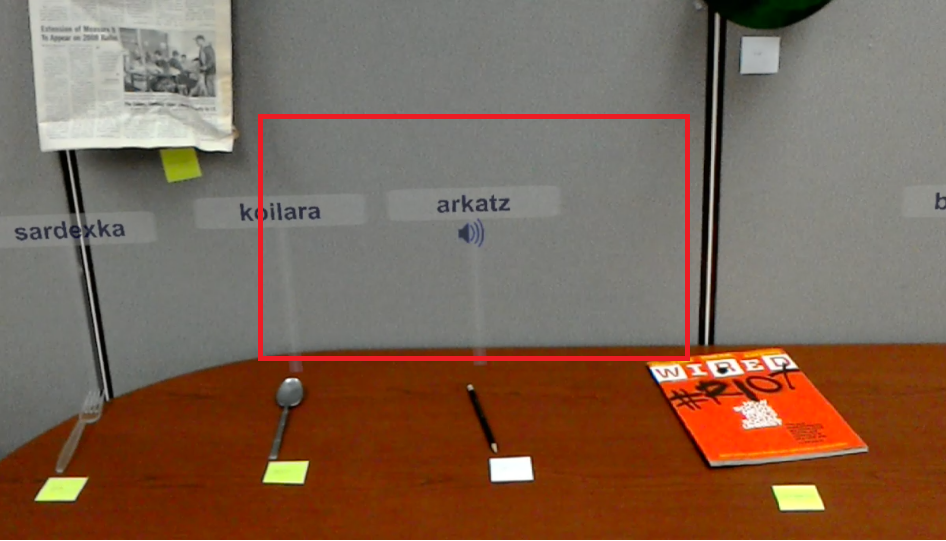}
  \caption{Example of the Basque labels shown in the AR condition of the experiment. This is a HoloLens {\it mixed reality capture} of the scene, which exaggerates the AR field of view. The approximate actual field of view for the label annotations is highlighted in red. }~\label{fig:ar-labels}
\end{figure}

The augmented reality modality (shown in Figure \ref{fig:ar-labels}) made use of a Microsoft HoloLens, an augmented reality head-mounted display. The application was set up in a room containing all of the objects from the two word groups, but only allowed the participants to see labels annotating the objects from the currently chosen subgroups with the relevant words in the foreign language. The device's real-time mapping of the room let the users walk around the room while keeping the labels in place, and save the location of the labels throughout the study, between users and after restarting the device. As a precaution, before every learning phase on the HoloLens, the administrators of the study verified that the labels were in place, and after handing over the device to the participants, that they were able to see every label. The app had two modes: "admin mode", allowing the instructor to place labels with voice commands or gestures, select which word subgroups to display, or enter a user ID; and a "user" mode that restricted these functionalities but allowed the participants to interact with labels during the learning task. 

On the HoloLens, the cursor's position is natively determined by the user's head orientation; in the app, moving the blue circle used as a cursor close to a label would turn the cursor into a speaker, signalling to the user the possibility to click to hear a recording of the word being pronounced through the device's embedded speakers. Each label had an extended, invisible hitbox to allow the users to click the labels more comfortably. Moreover, the labels' and hitboxes' sizes, along with the real objects' locations, were adjusted based on the room's dimensions and the device's field of view to ensure that the participants could not see more than two labels at the same time, and that looking at a label would most likely lead to the cursor being in that label's hitbox. This was used to log the attention given to each word during the learning task, in "user mode". 

In between the learning phases, "admin mode" was switched on to display a new subgroup, check on the labels, and temporarily disable logging of attention data. 

Due to the HoloLens's novelty, the participants were allowed to interact with animated holograms for as long as they wished before the AR learning task to get used to the controls, adjust the device and overcome some of the novelty factor of the modality.

\subsection{Learning Task}
The Basque language was chosen after ruling out numerous languages that shared too many cognates (words that can be recognised due to sharing roots with other languages) with English, Spanish and other languages that are commonly spoken in the region where the study was administered. Basque presented interesting properties: Latin alphabet to facilitate the learning, but generally regarded as a language isolate from the other commonly spoken languages \cite{trask1997history}, allowing us to control the number of cognates more easily, with one of the authors being fluent in the language. The 30 words were carefully chosen and split into two groups A and B based on difficulty and length, and further split into 3 subgroups per word group where each subgroup corresponded to a topic: A1 was composed of office related words (pen, pencil, paper, clock, notebook), A2 of kitchen related words (fork, spoon, cup, coffee, water), A3 of clothing related words (hat, socks, shirt, belt, glove), B1 of some other office related words (table, chair, scissors, cellphone, keyboard), B2 of printed items (newspaper, book, magazine, picture, calendar), and B3 of means of locomotion (car, airplane, train, rocket, horse). The study's counterbalancing helped address possible issues arising from A and B potentially not being balanced enough.
The learning task on a device was constituted of 3 learning phases, each of which lasted 90 seconds, for a total of 2 learning tasks (one per device) or 6 learning phases across both devices. The limit of 90 seconds was adjusted down from 180 seconds after a pilot study had shown a large ceiling effect with the users reporting having too much time. Once A or B was chosen as a group of words, the users successively saw subgroup 1 (5 words) during the first learning phase, then subgroups 1 and 2 (10 words) during the second learning phase, and then 1, 2 and 3 (15 words) in the last learning phase. The decision to allow the users to review the previous subgroups came as a solution to avoid the floor effects in the productive recall tests observed in the pilot study.

\subsection{Distraction Task}
In order to prevent the users from going straight from learning to testing, a distraction was used to reduce the risk of measuring only very short-term recall. The task needed to have enough cognitive load to distract the participants from the words they had just learnt. The participants' performance at the task should also be correlated to their general performance regarding the study, in order to avoid introducing new effects -- for example, a mathematical computation may bias the results as a participant with above average computational skills but below average memory skills may pass it fast enough that they would perform as well as another participant with below average computational skills but above average memory skills. Therefore, the distraction was chosen to be a memorisation task, in which the participants were asked to learn a different alphanumeric string of length 8 before every recognition test. The six codes used were the same for everyone, and were presented in the same order for every participant for the 2x2 balancing to mitigate possible ordering effects.


\section{Metrics}
To best understand the lasting impact of learning in the different modalities, our metrics included production and recognition of vocabulary, both immediately after a learning session, and also in test several days afterwards. 

\begin{figure}[t]
  \centering
  \includegraphics[width=3.3in]{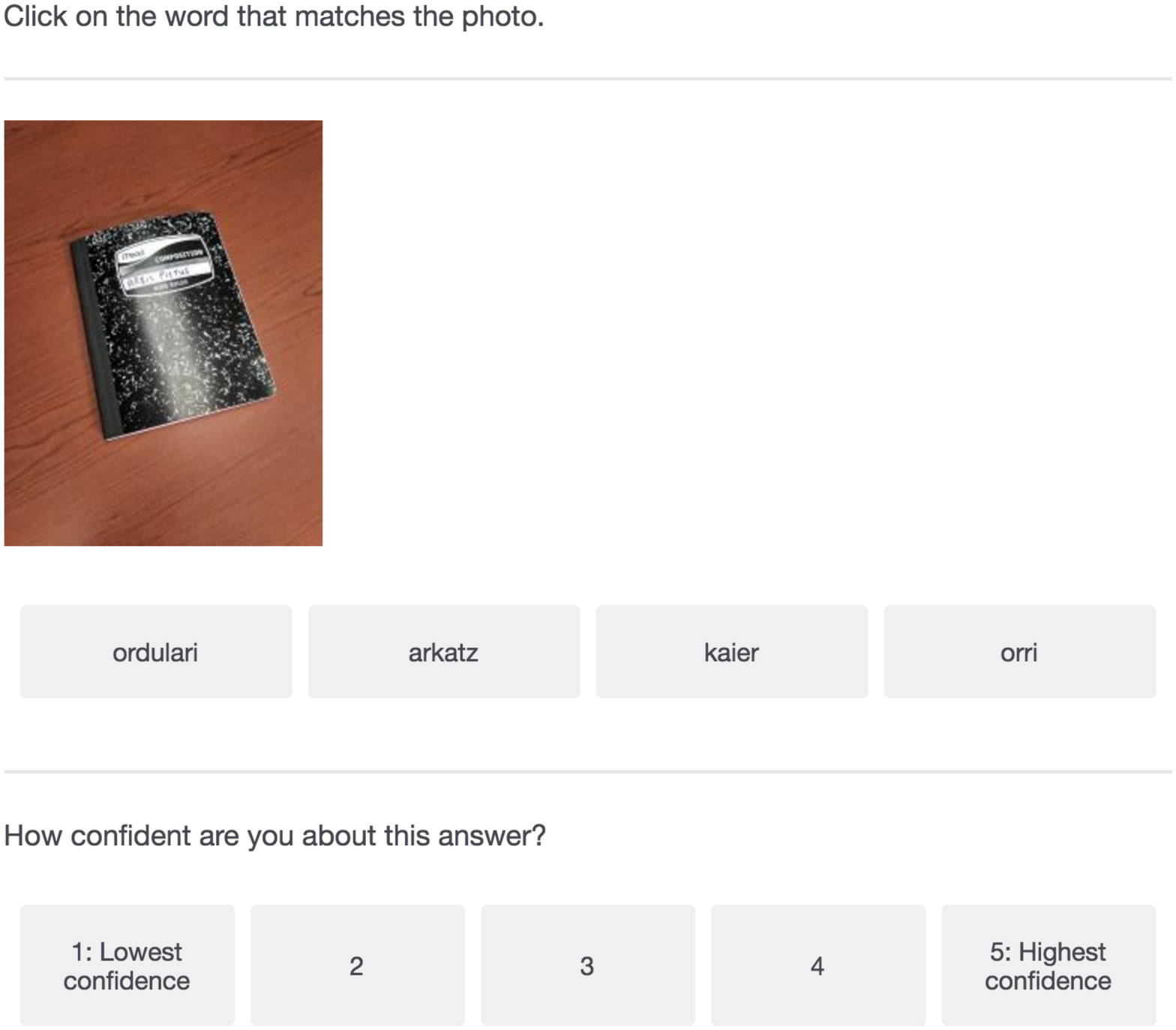}
  \caption{Format of the Productive Recognition Test. }~\label{fig:receptive-format}
\end{figure}

\subsection{Productive Recognition Test}

The productive recognition tests were administered on the desktop computer used for the questionnaire, in a different room from the two used for the learning tasks. Figure \ref{fig:receptive-format} shows the format of the test. The questions consisted in 5 images, each accompanied by a choice of 4 words from which the participants had to pick the appropriate one. Each image corresponded to one of the 5 new words seen in the preceding learning phase: A1 or B1 after the first learning phase, A2 or B2 after the second learning phase, and A3 or B3 after the third learning phase, depending on which one of A or B was chosen as the word group for that learning task, for a total of 6 recognition tests across the 2 learning tasks. All 5 images were available on the same page, allowing the participants to proceed by elimination. There was no time constraint, to avoid frustrating the participants, who were encouraged to use the tests as a way to prepare for the productive tests due to the strong floor effects observed in the pilot study. The performance was measured as either 1 for a correct answer, or 0 for an incorrect answer. Every question was accompanied by a confidence prompt on a scale of 5 ranging from "Lowest Confidence" to "Highest Confidence".

\subsection{Productive Recall Test}

\begin{figure}
  \centering
  \includegraphics[width=3.3in]{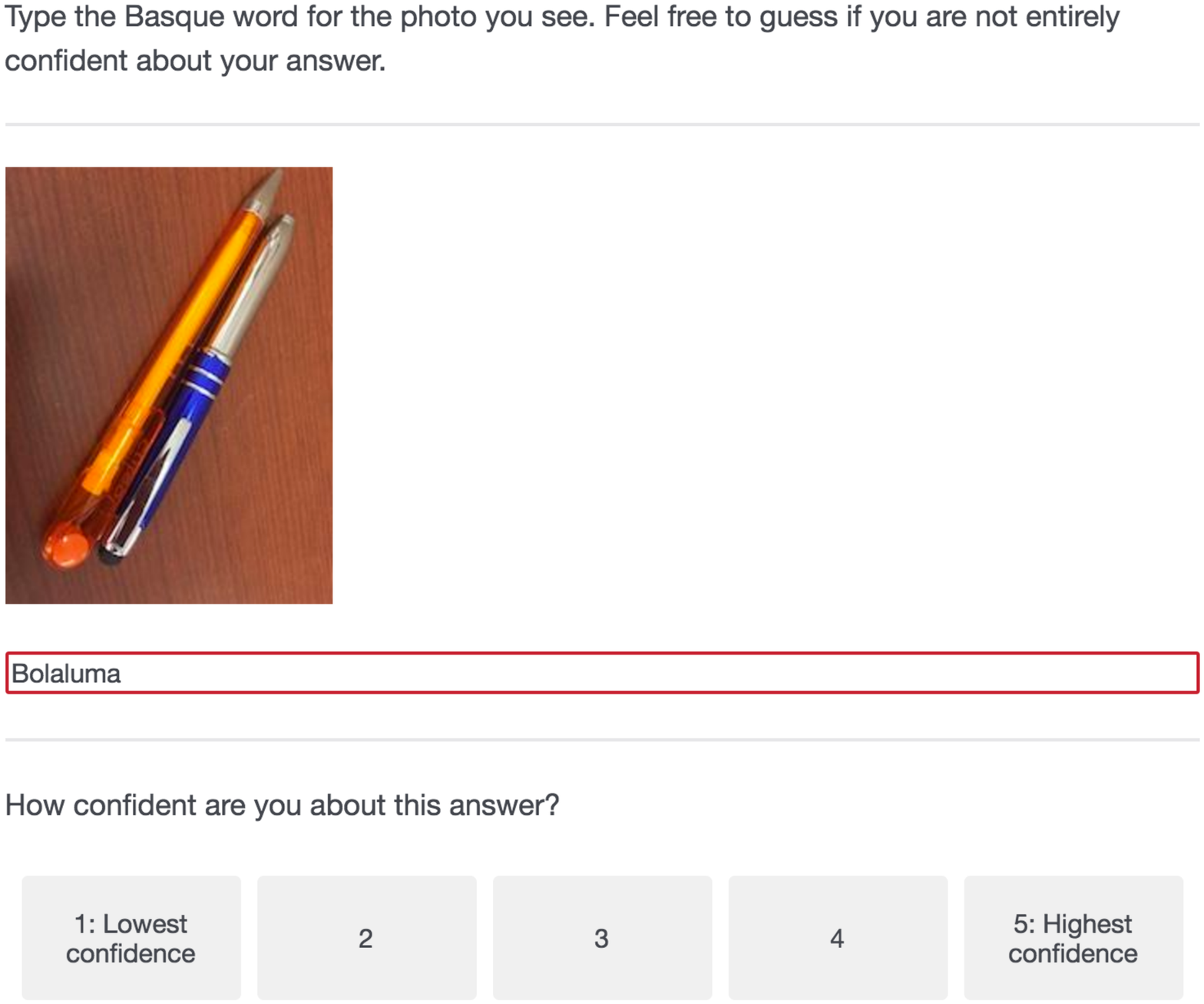}
  \caption{Format of the Productive Recall Test. }~\label{fig:productive-format}
\end{figure}

The productive recall tests took place on the same computer used for the recognition tests, immediately after the third recognition test at the end of each learning task.  Figure \ref{fig:productive-format} shows the format of the test, which also required a confidence evaluation for each answer.  The productive recall test had 15 images corresponding to the 15 words from the selected word group, and participants were asked to type the corresponding word in Basque below each image. The error on a participant's answer was measured using the Levenshtein distance, which counts the minimum number of insertions, deletions and substitutions needed to transform a word into another, between their answers and the correct spellings\cite{levenshtein1966binary}. Participants were therefore encouraged to try their best guess to get partial credit if they did not know the answer, and had to provide an answer to every question to end the test. The Levenshtein distance was also upper bounded in our analysis by the length of the (correctly spelled) word considered, to prevent answers such as "I don't remember" from biasing a participant's average error, and divided by the length of the correct answer to get a normalised error:
\begin{equation}
    \mathrm{AdjLev}(w, \hat{w}) = \frac{\mathrm{min}(\mathrm{Lev}(w, \hat{w}), \mathrm{Length}(\hat{w}))}{\mathrm{Length}(\hat{w})}
\end{equation}
where $w$ is the participant's answer on a given question, and $\hat{w}$ the correct answer. The score was then computed as
\begin{equation} \label{eq:score}
    \mathrm{Score}(w, \hat{w}) = 1 - \mathrm{AdjLev}(w, \hat{w})
\end{equation}
where 1 indicates a perfect spelling, and 0 a maximally incorrect answer. As in the recognition tests, every question was accompanied by a confidence prompt on a scale of 5 ranging from "Lowest Confidence" to "Highest Confidence".

\subsection{Delayed Test}
The delayed tests consisted of the same tests used for the same-day testing, in a slightly different order: the productive recall test of each word group was administered before the 3 recognition tests to prevent participants from reviewing with the recognition tests due to the absence of a time constraint. The tests were sent in a personalized email to the participants 4 days after the study. Only tests completed in the 24 hours after receiving the email were kept in the analysis. Further, the test did not allow the participants to press the back button, and only tests completed in a similar amount of time as the same-day tests were kept. Participants were informed that the study being comparative, the absolute number of words they remembered did not matter, and that the goal of the study was to measure how many people performed better with either device with no expectation of a modality being better than another. This was done in order to reduce the impact of potential demand effects such as bias towards either modality, and only their remembrance of the words (i.e. no qualitative feedback) was evaluated in the delayed test to further diminish such biases. In total, 31 participants' delayed test answers satisfied the criteria mentioned above. Note that the 2x2 counterbalancing was conserved (8 participants had followed order I, II and IV and 7 order III as defined in Table \ref{tab:ordering}).

\begin{figure*}[t]
  \centering
  \includegraphics[width=4in]{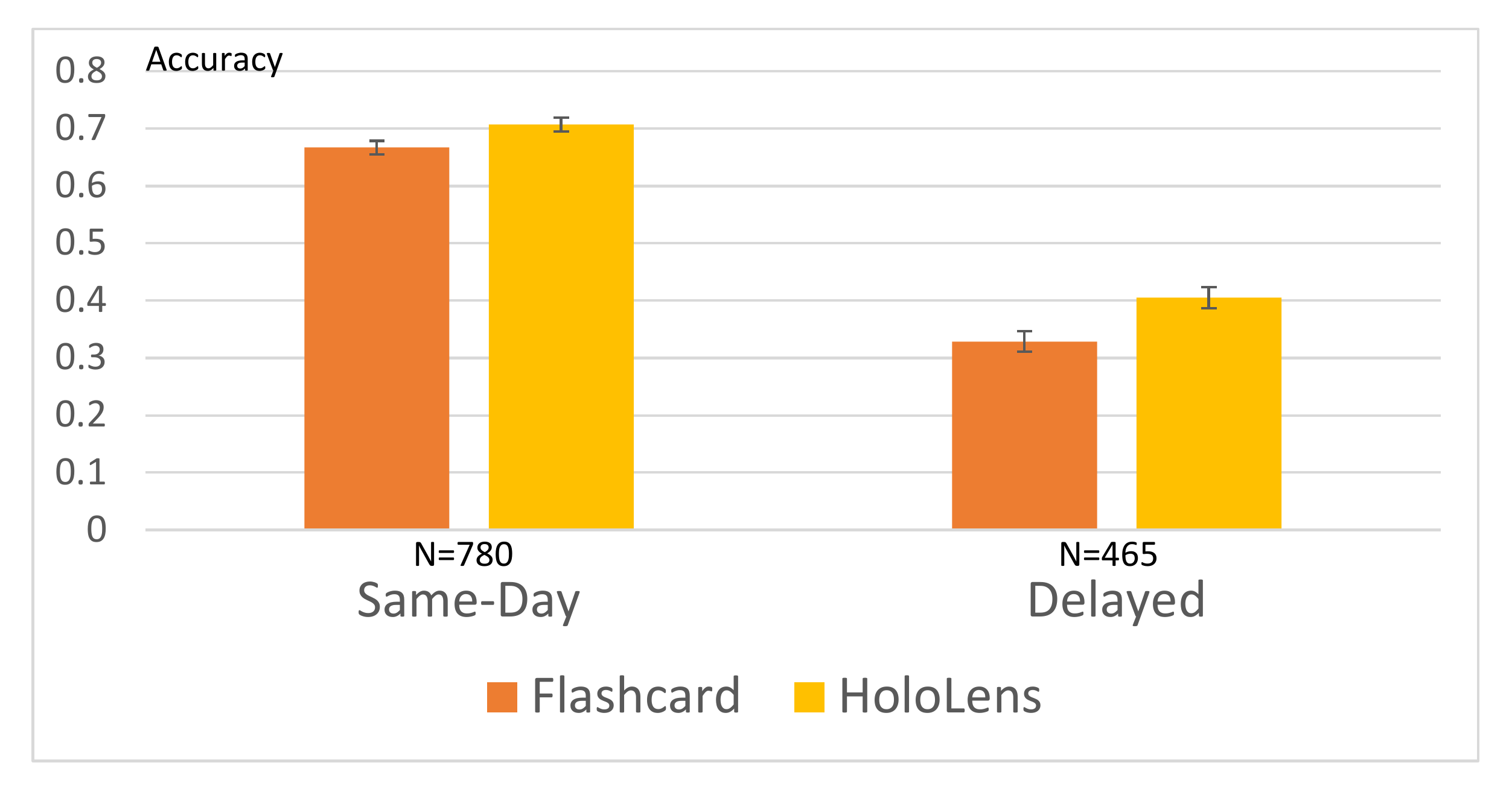}
  \caption{User performance on same-day and 4-day-delayed productive recall tests. The left group shows the flashcard and AR accuracy score for the same-day test and the right side shows the comparison for the 4-day-delayed test. Error bars show standard error here.  }~\label{fig:prod-recall}
\end{figure*}

\begin{figure*}[t]
  \centering
  \includegraphics[width=5in]{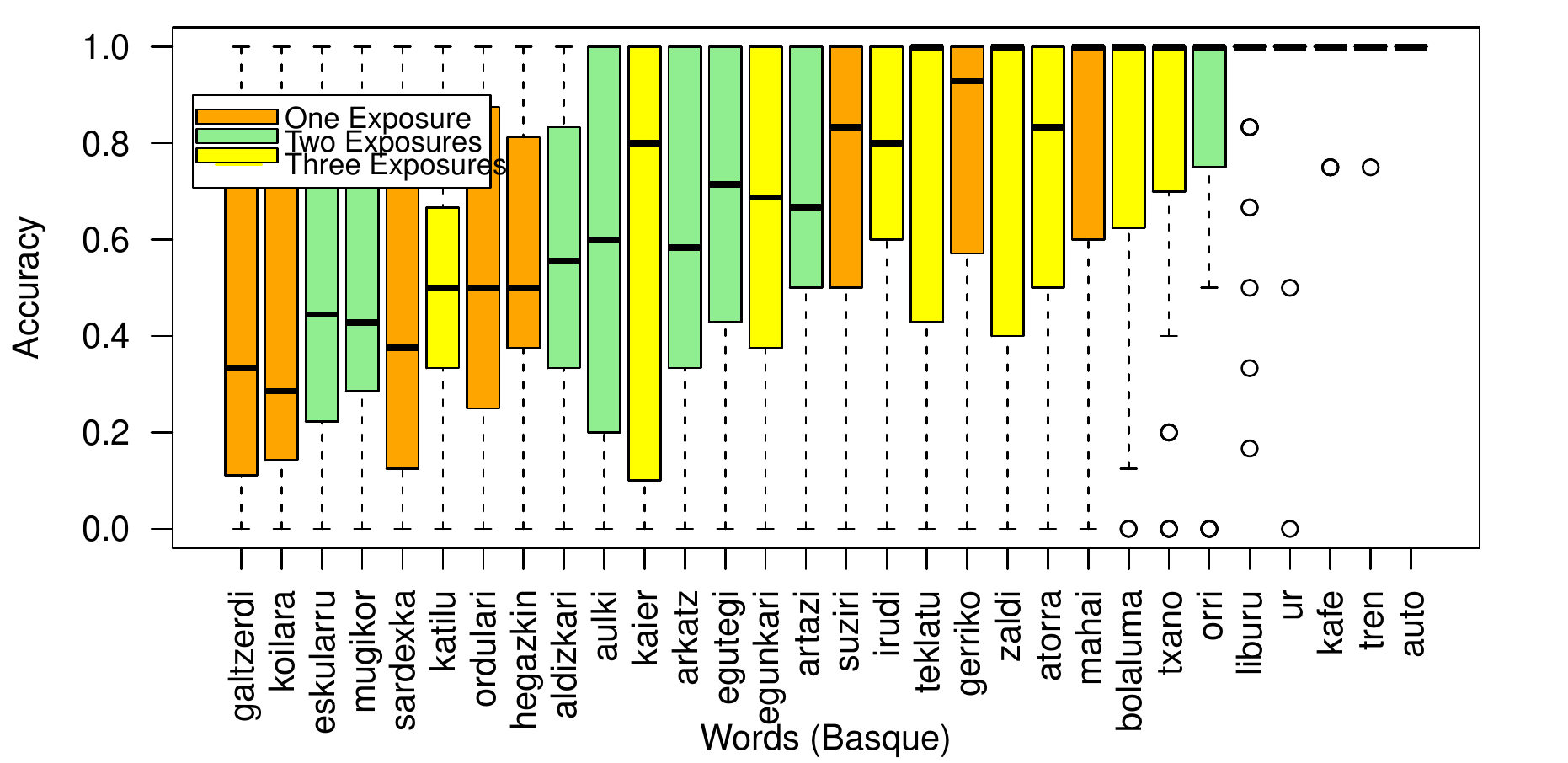}
  \caption{User performance on delayed productive recall tests, ranked by term. Colors show exposure groups.  The accuracy score on the y-axis is computed from the mean of the normalized Levenshtein distance between the participants' spelling and the correct spelling. }~\label{fig:delayed-prod-recall-byterm}
\end{figure*}

\section{Results}
Now that we have described our experimental setup and metrics, we present a discussion of results, organized by test type (recall and recognition), a brief discussion of attention metrics (gaze and click behavior), and a discussion of qualitative feedback from post-study questions and interviews.

\begin{table}[b]
\centering
\small
\label{tab:stat-quant}
\begin{tabular}{@{}llll@{}}
Dependent Variable (Accuracy) & Z       & \textit{p}     & effect size \\ \midrule
\textbf{Same-day Prod Recall}   & -2.5397 & \textbf{0.01109}  & 0.352       \\
\textbf{Delayed Prod Recall}   & -3.1959 & \textbf{0.001394} & 0.574       \\
Same-day Recognition  & -0.7926 & 0.42799           & 0.110       \\
Delayed Recognition  & -0.1239 & 0.901389          & 0.022       \\
Same-day Prod Recall (FC pref group)    & 1.1589  & 0.246488          & 0.237       \\
Delayed Prod Recall (FC pref group)    & -0.0580 & 0.95367           & 0.016      
\end{tabular}
\caption{Key results from statistical analysis. Results highlighted in bold face are statistically significant effects.}
\end{table}

\subsection{Productive Recall}\label{sec:prod}

Figure \ref{fig:prod-recall} shows the accuracy results of the same-day productive recall test compared to the delayed test for both modalities. The AR condition is shown in the lighter color.  The delayed test was administered 4 days after the main study, and there was some attrition, with 780 question responses in the main study and 465 for the delayed.  Accuracy was measured using the score function previously defined in Eq.\ref{eq:score} as 1 minus the normalized Levenshtein distance between the attempted spelling and the correct spelling.  In the same-day test, the AR condition outperformed the flashcards condition by 7\%, and more interestingly, in the delayed test, this improvement was more pronounced, at 21\% better than the flashcard condition. The test results were analyzed in a non-parametric way after Shapiro-Wilk tests confirmed the non-normality of the data. This is due in part to the many occurrences of words perfectly spelled. Both differences are significant with Wilcoxon Signed-rank tests: p=0.011 and p=0.001 for the same-day and the delayed productive results respectively, as seen in Table 2.  The table also reports productive recall scores for those users who reported that Flashcards were more effective than AR (FC pref Group).  Interestingly, no significant difference was found between the modalities for this sub-group, in contrast to the results for the general population.

Based on interviews with the participants, we believe that the significant improvement in delayed recall is linked to the spatial aspect in the HoloLens condition. Several participants reported qualitative feedback to this effect, such as in the following example:
\textit{``One reason the AR headset helped me recognize the words better is because of the position of the object. Sometimes, I'm not memorizing the word, I'm just recognizing the position of the object and which word it correlates to. ''}

\subsection{Productive Recognition}
Productive recognition was analyzed in the same manner as productive recall, however a histogram of response accuracy revealed a ceiling effect in the data, where many participants provided fully correct responses.  The mean productive recognition score was 0.89 for the same-day test in both modalities and 0.84 in the 4-day delayed test again for both modalities.  In the delayed test, the productive recall was presented first to avoid learning effects from viewing multiple choice options.  There was no significant difference between the modalities in this test.

\subsection{Attention Metrics}
Gaze data was gathered for both modalities. For the flash card application, we collected eye tracking data using a screen-based eye tracker, and for the HoloLens application we recorded head orientation focus as described above. 

For each of the terms in the three different exposure groups we computed the average time that participants' attention was focused on that item.  This was performed primarily to examine why repeated exposure to terms did not produce an observed improvement in accuracy. For the first group, the mean was 13.5 seconds (SD 6.3 seconds), for the second, the mean was 10.8 seconds (SD 7.2 seconds), and third group had a mean attention time of 7.2 seconds (SD 4.5 seconds). The differences in attention times not being significant between the different groups may imply that during the learning phases, participants focused mainly on the new items, or that users chose to focus on different words on average. This is reinforced by the fact that no significant accuracy improvement was measured for repeated-exposure items, as evidenced by Figure \ref{fig:delayed-prod-recall-byterm} where the most significant effect is word length.  \\
Click data was recorded for the flashcard application to help identify potential learning patterns. Recall that the flashcards had two sides and required a click to turn from text to image and back again (Figure \ref{fig:flash}). The click patterns showed that people tended to click more often towards the end of the study. 18 of the participants had a pattern of clicking the same flashcard over 5 times in a row, perhaps indicating a desire to see both image and text at the same time, or testing themselves during the learning phase. Both possibilities are supported by users reporting in the post-study interview that they enjoyed the ability to see the object and the word simultaneously in AR, while others mentioned making use of the flashcards' two-sided nature to self-test. 

\subsection{Perception and Qualitative Feedback}
Participants were asked about their experience using AR and flashcards, and their subjective ratings correspond with their learning performance. In terms of what was fastest for learning words, 54\% found AR fastest, compared to 46\% who found flashcards fastest. As a side note, 13 among the delayed test population had reported preferring the flashcards, as opposed to 18 for AR. As for the learning experience, 75\% of participants rated AR ``good'' or ``excellent'', while 63\% rated flashcards ``good'' or ``excellent''.

\begin{figure*}
  \centering
  \includegraphics[width=6.0in]{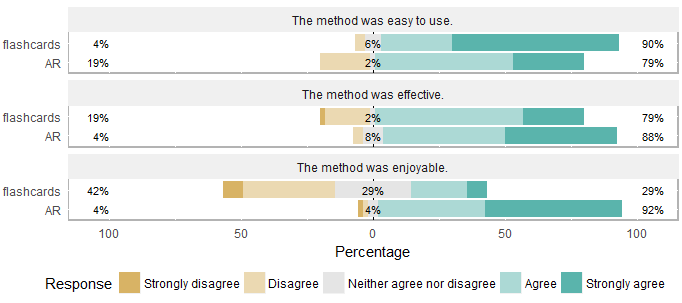}
  \caption{Qualitative feedback from the 52 participants}
  \label{fig:qual-feedback}
\end{figure*}

Figure \ref{fig:qual-feedback} shows that when asked about the effectiveness of each platform for learning words, 88\% of participants ``somewhat agreed'' or ``strongly agreed'' that the AR headset was effective, while 79\% ``somewhat agreed'' or ``strongly agreed'' that the flashcards were effective.

Participants' comments comparing the two platforms revealed that about 20\% (10 of 52) felt AR and flashcards were equally effective for learning because of the visual imagery both provide. 14 of 52 specifically mentioned that they found AR better because they saw the word and object at the same time. Almost 20\% (10 of 52) stated that AR was better because it was more interactive, immersive, and showed objects in real time and space (e.g., \textit{"The flashcards are classic and I have experience learning from them but the AR headset was more immersive"} and \textit{"The headset was more interactive because it was right in front of you with physical objects rather than through a computer screen"}). Only 13\% of the participants commented that flashcards were better, due to their familiarity with similar apps and computers in general.

A stark/striking difference was found in participants' opinions about which platform was enjoyable for learning. Figure \ref{fig:qual-feedback} shows that 92\% of participants ``somewhat agreed'' or ``strongly agreed'' that using the AR headset was enjoyable for learning words, compared to only 29\% for using the flashcards. Open-ended comments from the participants pointed to the not unexpected novelty effect of AR (21 of 52 or 40\%), \textit{"The AR Headset because it was an incredibly futuristic experience."} In addition, 16 of 52 participants (31\%) commented explicitly on how AR is more interactive, engaging, hands-on, natural, and allowed for physical movement (e.g., \textit{``The AR headset was more interactive and required movement which engaged my mind more''} and \textit{``The AR Headset was more fun because it's more fun to be able to move around and see things in actual space than on a computer screen''} or \textit{``The AR headset was more enjoyable because it allowed for you to interact with the objects that you are learning about. It felt more realistic and applicable to real life, plus I had the visual image that helped me remember the words''}). Only 8 of 52 participants (15\%) indicated that flashcards were more enjoyable because they were familiar, practical, and straightforward.

As we noted earlier in the discussion of productive recall results (Section \ref{sec:prod}, several participants commented in interviews or left text feedback related to the spatial aspect of the AR condition, generally saying that it helped give them an extra dimension to aid in learning. For example, one participant reported that: 
\textit{``The AR headset put me in contact with the objects as well as had me move around to find words. I was able to recall what words meant by referencing their position in the room or proximity to other objects as well. ***Seeing the object at the same time as the word strengthened the association for me greatly***''}.  Another participant said \textit{
``the AR seems like it would work better with friends or family trying to learn together, while the flashcards seem to work on an individual level.''}. The latter comment points towards a social or interactive aspect of AR-based learning which we have not focused on in this study, but is nonetheless of potential interest to system designers and language learning researchers. The potential for social interaction and learning that this participant mentioned is likely linked to the availability of an interactive learning space.

Another possible benefit to learning in the AR condition is that it can facilitate the so called "method of loci" or "memory palace" technique \cite{Yates1966}. It has been shown to be useful when applied to learn the vocabulary of a foreign language. The method is described for example in \cite{metivier2012learn}. The author suggests to begin by creating a memory palace for each letter of the German alphabet by associating it with a location in an imaginary physical space. Each memory palace then is recommended to include a number of loci where an entry (a word or a phrase) can be stored and recalled whenever it is needed.  One of our participants made a comment about this learning method after learning in the AR condition: \textit{``I use memory palaces, so I really enjoyed AR as it felt somewhat familiar and made it easier for me to use the technique than the flashcards''.}


\section{Limitations and Future Work}

Our proof-of-concept experiment shows that AR can produce better results on the learning of foreign-language nouns in a controlled lab-based user study. However, the study has several limitations. First, learning itself occurred in a controlled experimental context, in which subjects were paid an incentive.  This cannot be assumed to be representative of real-world learning, and it is possible that our results may vary in real learning contexts.  Second, and related, it is likely that novelty effects had some impact on the study given that the HoloLens remains in the category of new and exciting technology. Our design included a long acclimatisation phase with the device, but it is difficult to be sure that our qualitative results have not been impacted by novelty effects. Third, we chose to adopt a simple and standard implementation of the flashcard system, with words and pictures on opposite sides. This was fundamentally different from the AR condition, wherein labels and objects were visible at the same time.  A small number of participants noted that they preferred the ability to view the object label and the object at the same time in the AR condition. On the other hand, others made use of the ability to self test in the flashcard condition.  We are aware that our design choices and trade-offs will impact the learning experience. It is possible that other implementations would produce different results.  Our results here represent a black-box comparison of these two learning approaches, and we encourage other researchers to extend our study to include other learning platforms and designs. Last, our productive recognition tests, while carefully controlled based on informal pre-studies and performance information from existing literature, showed ceiling effects with a large number of participants. No ceiling or floor effects were observed for the productive recall test.  In follow-up experiments, we will increase the difficulty of the productive recognition tests. 

There are several avenues to continue our research on the ARbis Pictus system, most notably, by taking the system beyond the controlled learning environment that was described in this paper and applying it to real-world learning tasks. As an initial step towards this, we have implemented a first-draft of real time object labeling with HoloLens and YOLO \cite{redmon2016yolo9000} and are also in the process of working with students and course administrators to develop personalized language learning plans that could be deployed in the system.  Evaluating the performance of a real-world AR personalized-learning system is clearly a non-trivial task that will require complex longitudinal studies with many learners to account for differences in user experiences brought about by uncontrolled data in real-world environments. Ideally, we would like to lead such a longitudinal study over the course of several weeks in a classroom, in order to measure the potential of AR in less controlled environments where the influence of novelty may be easier to measure and where students may interact with each others. In terms of education and learning theory, it may be possible for our results to contribute to expanding the existing and established theories of CTML. We also intend on expanding this work with more analysis of how various groups, such as multilingual people, approach vocabulary learning, as indicated by interaction patterns, gaze data and performance.


\section{Conclusion}

This paper has described a 2x2 within-subjects experimental evaluation (N=52) to assess the effect of AR on learning of foreign-language nouns compared to a traditional flashcard approach.  Key research questions were proposed, related to quantitative performance in immediate and delayed recall tests, and user experience with the learning modality (qualitative data).  Results show that 1.) AR outperforms flashcards on productive recall tests administered same-day by 7\% (Wilcoxon Signed-rank p=0.011), and this difference increases to 21\% (p=0.001) in productive recall tests administered 4 days later.  2.) Participants reported that the AR learning experience was both more effective and more enjoyable than the flashcard approach.  We believe that this is a good indication that AR can be beneficial for language learning, and we hope it may inspire HCI and education researchers to conduct comparative studies. 


\acknowledgments{
Parts of this research were supported by ONR grant \#N00014-16-1-3002 and NSF DRL planning grant \#1427729. Thanks to Matthew Turk, Yun Suk Chang, and JB Lanier for their contributions and feedback on the project.}

\bibliographystyle{abbrv-doi}

\bibliography{template}

\begin{thebibliography}{10}

\bibitem{cai2014wait}
C.~J. Cai, P.~J. Guo, J.~Glass, and R.~C. Miller.
\newblock Wait-learning: leveraging conversational dead time for second
  language education.
\newblock In {\em CHI'14 Extended Abstracts on Human Factors in Computing
  Systems}, pp. 2239--2244. ACM, 2014.

\bibitem{chun1996effects}
D.~M. Chun and J.~L. Plass.
\newblock Effects of multimedia annotations on vocabulary acquisition.
\newblock {\em The modern language journal}, 80(2):183--198, 1996.

\bibitem{costabile2008explore}
M.~F. Costabile, A.~De~Angeli, R.~Lanzilotti, C.~Ardito, P.~Buono, and
  T.~Pederson.
\newblock Explore! possibilities and challenges of mobile learning.
\newblock In {\em Proceedings of the SIGCHI Conference on Human Factors in
  Computing Systems}, pp. 145--154. ACM, 2008.

\bibitem{culbertson2016social}
G.~Culbertson, S.~Wang, M.~Jung, and E.~Andersen.
\newblock Social situational language learning through an online 3d game.
\newblock In {\em Proceedings of the 2016 CHI Conference on Human Factors in
  Computing Systems}, pp. 957--968. ACM, 2016.

\bibitem{dunleavy2014augmented}
M.~Dunleavy and C.~Dede.
\newblock Augmented reality teaching and learning.
\newblock In {\em Handbook of research on educational communications and
  technology}, pp. 735--745. Springer, 2014.

\bibitem{Fujimoto2012a}
Y.~Fujimoto, G.~Yamamoto, T.~Taketomi, J.~Miyazaki, and H.~Kato.
\newblock {Relationship between features of augmented reality and user
  memorization}.
\newblock In {\em 2012 IEEE International Symposium on Mixed and Augmented
  Reality (ISMAR)}, pp. 279--280. IEEE, Nov. 2012.

\bibitem{godwin2016augmented}
R.~Godwin-Jones.
\newblock Augmented reality and language learning: From annotated vocabulary to
  place-based mobile games.
\newblock {\em Language learning and technology}, 20(3):9--19, 2016.

\bibitem{grasset2007mixed}
R.~Grasset, A.~Duenser, H.~Seichter, and M.~Billinghurst.
\newblock The mixed reality book: a new multimedia reading experience.
\newblock In {\em CHI'07 extended abstracts on Human factors in computing
  systems}, pp. 1953--1958. ACM, 2007.

\bibitem{Ishikawa2008}
T.~Ishikawa, H.~Fujiwara, O.~Imai, and A.~Okabe.
\newblock {Wayfinding with a GPS-based mobile navigation system: A comparison
  with maps and direct experience}.
\newblock {\em Journal of Environmental Psychology}, 28(1):74--82, mar 2008.

\bibitem{kaufmann2003mathematics}
H.~Kaufmann and D.~Schmalstieg.
\newblock Mathematics and geometry education with collaborative augmented
  reality.
\newblock {\em Computers \& graphics}, 27(3):339--345, 2003.

\bibitem{levenshtein1966binary}
V.~I. Levenshtein.
\newblock Binary codes capable of correcting deletions, insertions, and
  reversals.
\newblock In {\em Soviet physics doklady}, vol.~10, pp. 707--710, 1966.

\bibitem{liu2016analyzing}
Y.~Liu, D.~Holden, and D.~Zheng.
\newblock Analyzing students' language learning experience in an augmented
  reality mobile game: an exploration of an emergent learning environment.
\newblock {\em Procedia-Social and Behavioral Sciences}, 228:369--374, 2016.

\bibitem{mayer2005cambridge}
R.~E. Mayer.
\newblock {\em The Cambridge handbook of multimedia learning}.
\newblock Cambridge University Press, 2005.

\bibitem{mayer2011applying}
R.~E. Mayer.
\newblock {\em Applying the science of learning}.
\newblock Pearson/Allyn \& Bacon Boston, MA, 2011.

\bibitem{mayer1994whom}
R.~E. Mayer and V.~K. Sims.
\newblock For whom is a picture worth a thousand words? extensions of a
  dual-coding theory of multimedia learning.
\newblock {\em Journal of educational psychology}, 86(3):389, 1994.

\bibitem{metivier2012learn}
A.~Metivier.
\newblock {\em How to Learn and Memorize German Vocabulary: ... Using a Memory
  Palace Specifically Designed for the German Language (and Adaptable to Many
  Other Languages Too)}.
\newblock CreateSpace Independent Publishing Platform, 2012.

\bibitem{moreno1999cognitive}
R.~Moreno and R.~E. Mayer.
\newblock Cognitive principles of multimedia learning: The role of modality and
  contiguity.
\newblock {\em Journal of educational psychology}, 91(2):358, 1999.

\bibitem{nakata2011computer}
T.~Nakata.
\newblock Computer-assisted second language vocabulary learning in a
  paired-associate paradigm: A critical investigation of flashcard software.
\newblock {\em Computer Assisted Language Learning}, 24(1):17--38, 2011.

\bibitem{plass1998supporting}
J.~L. Plass, D.~M. Chun, R.~E. Mayer, and D.~Leutner.
\newblock Supporting visual and verbal learning preferences in a
  second-language multimedia learning environment.
\newblock {\em Journal of educational psychology}, 90(1):25--36, 1998.

\bibitem{redmon2016yolo9000}
J.~Redmon and A.~Farhadi.
\newblock Yolo9000: Better, faster, stronger.
\newblock {\em arXiv preprint arXiv:1612.08242}, 2016.

\bibitem{Rosello2016}
O.~Rosello, M.~Exposito, and P.~Maes.
\newblock {NeverMind: Using Augmented Reality for Memorization}.
\newblock In {\em Proceedings of the 29th Annual Symposium on User Interface
  Software and Technology - UIST '16 Adjunct}, pp. 215--216. ACM Press, New
  York, New York, USA, 2016.

\bibitem{schmalstieg2007experiences}
D.~Schmalstieg and D.~Wagner.
\newblock Experiences with handheld augmented reality.
\newblock In {\em Mixed and Augmented Reality, 2007. ISMAR 2007. 6th IEEE and
  ACM International Symposium on}, pp. 3--18. IEEE, 2007.

\bibitem{scrivner2016augmented}
O.~Scrivner, J.~Madewell, C.~Buckley, and N.~Perez.
\newblock Augmented reality digital technologies (ardt) for foreign language
  teaching and learning.
\newblock In {\em Future Technologies Conference (FTC)}, pp. 395--398. IEEE,
  2016.

\bibitem{seedhouse2014european}
P.~Seedhouse, A.~Preston, P.~Olivier, D.~Jackson, P.~Heslop, M.~Balaam,
  A.~Rafiev, and M.~Kipling.
\newblock The european digital kitchen project.
\newblock {\em Bellaterra journal of teaching and learning language and
  literature}, 7(1):0001--16, 2014.

\bibitem{trask1997history}
R.~Trask.
\newblock {\em The History of Basque}.
\newblock Routledge, 1997.

\bibitem{turk2014effects}
E.~T{\"u}rk and G.~Er{\c{c}}etin.
\newblock Effects of interactive versus simultaneous display of multimedia
  glosses on l2 reading comprehension and incidental vocabulary learning.
\newblock {\em Computer Assisted Language Learning}, 27(1):1--25, 2014.

\bibitem{yannier2015learning}
N.~Yannier, K.~R. Koedinger, and S.~E. Hudson.
\newblock Learning from mixed-reality games: Is shaking a tablet as effective
  as physical observation?
\newblock In {\em Proceedings of the 33rd Annual ACM Conference on Human
  Factors in Computing Systems}, pp. 1045--1054. ACM, 2015.

\bibitem{Yates1966}
F.~A. Yates.
\newblock {\em {The Art of Memory}}.
\newblock University of Chicago Press, 1966.

\bibitem{yoshii2006}
M.~Yoshii.
\newblock L1 and l2 glosses: Their effects on incidental vocabulary learning.
\newblock {\em Language learning and technology}, 10(3):85--101, 2006.

\end{thebibliography}
\end{document}